\documentclass[aps,pre,reprint,superscriptaddress,amsmath,amssymb,byrevtex,floatfix]{revtex4-2}

\usepackage{graphicx}
\usepackage{siunitx}
\usepackage[colorlinks=true, allcolors=blue]{hyperref}
\usepackage{amsmath}
\usepackage[normalem]{ulem}
\usepackage{mathtools}
\usepackage{bm}

\usepackage{color}
\definecolor{colortodo}{RGB}{255,0,0}

\definecolor{colortodo2}{RGB}{0,150,0}

\definecolor{colortodo3}{RGB}{0,0,255}

\begin{document}

\title{Emergence of tip singularities in dissolution patterns.}

\author{Martin Chaigne} \email{martin.chaigne@u-paris.fr} \affiliation{Mati\`ere et Syst\`emes Complexes (MSC), Université Paris Cité, CNRS (UMR 7057), 75013 Paris, France}
\author{Sabrina Carpy}  \affiliation{Laboratoire de Plan\'etologie et G\'eosciences (LPG), Nantes Université, CNRS (UMR 6112), 44322 Nantes, France}
\author{Marion Mass\'e} \affiliation{Laboratoire de Plan\'etologie et G\'eosciences (LPG), Nantes Université, CNRS (UMR 6112), 44322 Nantes, France}
\author{Julien Derr} \affiliation{Laboratoire Reproduction et D\'eveloppement des Plantes (RDP), Université de Lyon 1, ENS-Lyon, INRAE, CNRS, UCBL, 69364 Lyon, France}
\author{Sylvain Courrech du Pont} \affiliation{Mati\`ere et Syst\`emes Complexes (MSC), Université Paris Cité, CNRS (UMR 7057), 75013 Paris, France}
\author{Michael Berhanu} \affiliation{Mati\`ere et Syst\`emes Complexes (MSC), Université Paris Cité, CNRS (UMR 7057), 75013 Paris, France}

\date{\today}


\begin{abstract}
{Chemical erosion, one of the two major erosion processes along with mechanical erosion, occurs when a soluble rock-like salt, gypsum, or limestone is dissolved in contact with a water flow. The coupling between the geometry of the rocks, the mass transfer, and the flow leads to the formation of remarkable patterns, like scallop patterns in caves. We emphasize the common presence of very sharp shapes and spikes, despite the diversity of hydrodynamic conditions and the nature of the soluble materials. We explain the generic emergence of such spikes in dissolution processes by a geometrical approach. Singularities at the interface emerge as a consequence of the erosion directed in the normal direction, when the surface displays curvature variations, like those associated with a dissolution pattern. First, we demonstrate the presence of singular structures in natural interfaces shaped by dissolution. Then, we propose simple surface evolution models of increasing complexity demonstrating the emergence of spikes and allowing us to explain at long term by coarsening the formation of cellular structures. Finally, we perform a dissolution pattern experiment driven by solutal convection, and we report the emergence of a cellular pattern following well the model predictions. Although the precise prediction of dissolution shapes necessitates performing a complete hydrodynamic study, we show that the characteristic spikes which are reported ultimately for dissolution shapes are explained generically by geometrical arguments due to the surface evolution. These findings can be applied to other ablation patterns, reported for example in melting ice.}
\end{abstract}

\maketitle

Cusps, tips or pinches that can occur at the free surface of liquids have always attracted attention, notably because they often present self-similar shapes~\cite{Moffatt1992,Podgorski2001,Courrech2006,Lagarde2018}. Mathematically, the location where the interface curvature diverges is generically called a singularity~\cite{EggersBook}. 
But singular shapes can also occur on evolving solid surfaces carved by hydrodynamic processes, which is of prime importance in geomorphology.
At the surface of the Earth, landscapes are indeed shaped by erosion~\cite{Jerolmack2019}: water and wind carve rocks and mountains, dig valleys and caves, and sometimes produce spectacular structures. In mechanical erosion, sediments are ripped from the rock and carried by the flow. In chemical erosion, minerals dissolve into water before being carried as solutes~\cite{Meakin2010,Jamtveit2012}. In nature, this is the main erosion mechanism for limestone, gypsum or salt. When water flows over these rocks, it can lead to the formation of remarkable patterns. 
For example, sloping rain-exposed soluble ground can be covered by Rillenkarren (parallel grooves), in which wide concave channels are separated by narrow crests \cite{Lundberg2013,Guerin2020,Bertagni2021}. Even more impressive, stone forest of sharp pinnacles are observed on limestone in tropical Karst regions~\cite{FordWilliamsbook,Karrenbook}. Recent experimental, theoretical and numerical studies~\cite{Huang2020,Pegler2020,Pegler2021,Huang2022} report the similar formation of sharp pinnacles by dissolution of fast dissolving material like home-made candies, under the action of a gravity-driven convection flow, when the boundary layer charged in solute remains attached to the dissolving interface. 

Below the Earth surface, on the limestone walls of caves carved by underground rivers, scallop patterns are another important example of sharp structures generated by dissolution. The scallops appear as concave depressions rounded by sharp crests~\cite{Curl1966,FordWilliamsbook} with typical length scales ranging from centimeters to meters. They originate from a coupling between the topography and a flow which generates wall undulations that could be linear transverse (called ripples or flutes~\cite{Blumberg1974,Carpy2023}) or more complex (scallops)~\cite{Curl1966,Allen1970,Blumberg1974}. Recent modeling works~\cite{Claudin2017,Carpy2023} explains the emergence of the undulations as a linear instability mechanism at the laminar-turbulent transition, enhancing dissolution in the troughs. This two-dimensional mechanism predicts the wavelength as a function of the current velocity, but does not explain the sharp crests of the scallops and their evolution in the nonlinear regime of flow-dissolving surface interaction. Analogous scallop patterns have also been  generated in dissolution experiments without imposing flow, when the solute boundary layer detaches from the dissolving interface, generating a convection flow~\cite{Sullivan1996,Cohen2016,Cohen2020}. The roughness and stripes generated at short time by dissolution evolve into an assembly of concave troughs delimited by sharp crests~\cite{Cohen2020}. 

In all of these examples, the ultimate shapes result from different, specific and complex out of-equilibrium hydrodynamic problems, but generically exhibit sharp structures and spikes.\\

In this article, we show that a simple geometrical model of interface evolution predicts the emergence of tip singularities in finite time, which explains the generality of singular shapes on structures eroded by dissolution. First, we detect and characterize the presence of crests on a field example, a scalloped wall in the Saint-Marcel cave. Then, following the pioneering idea of A. Lange in the fifties~\cite{Lange1959,Curl1966}, we expose a model of interface evolution with a normal ablation leading to the emergence of singularities in finite time.  One-dimensional simulations show that an interface with initially non-uniform curvature evolves into an interface with gradient discontinuities, similar to the shocks emerging in interfaces following the deterministic Kardar-Parisi-Zhang (KPZ) equation~\cite{KardarPRL1986}. We also show that this observation is robust even when the erosion rate is non uniform, allowing patterns to emerge. In two dimensions, such a process generates, by coarsening, a cellular pattern. Finally, for the example of dissolution experiments driven by solutal convection, we analyze the emergence and the evolution of the scallop patterns in relation with the outcomes of our model.

\section{Characterisation of singularities in the field.}

\begin{figure*}[t]
   \centering
   \includegraphics[width=1\textwidth]{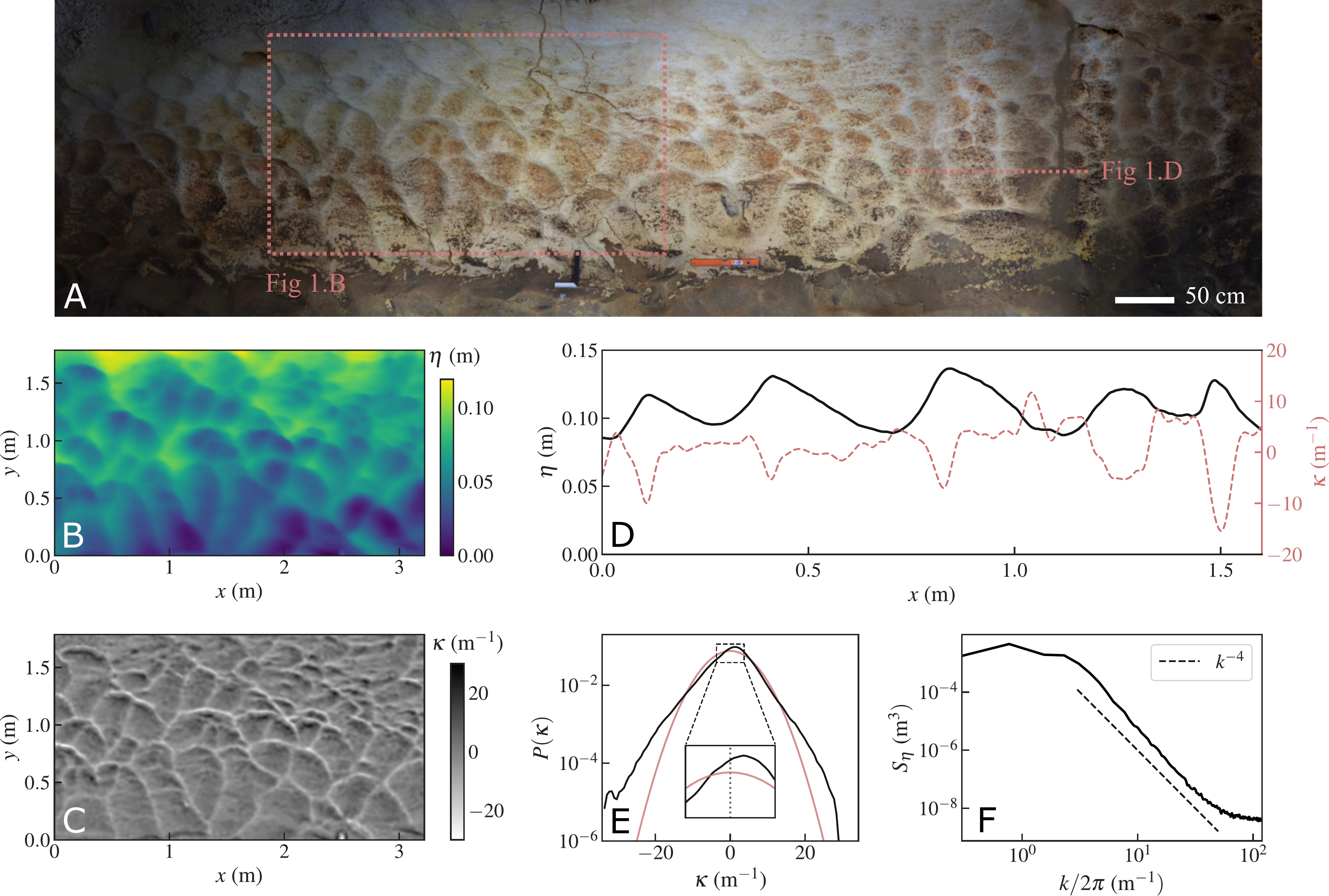}
   \caption{\label{SaintMarcel} Field measurements: scallops in the Saint Marcel cave (France). A: Orthophotograph of a vertical wall of the cave, approximately 2 meters high and 10 meters long, covered by scallops. B: 3D reconstruction of a portion of the wall (area surrounded by a pink dotted line on the orthophotograph) using photogrammetry. C: Mean curvature $\kappa$ of the topography for the same portion of the wall. Lines with highly negative curvature indicate location of crests. D: Longitudinal cross-section at the location of the pink dotted line on the orthophotograph (black solid line) and corresponding mean curvature (pink dashed line). E: Normalized mean curvature distribution. Solid pink line indicates Gaussian distribution with zero mean and same standard deviation. Inset: shape of the distribution around zero revealing a shift toward positive values. F: Power Spectral Density of the topography. Black dashed line corresponds to a power-law in $k^{-4}$.} 
\end{figure*}

The Saint-Marcel cave, located in the Ard\`eche department in southern France, is an impressive limestone cave open to the public, famous for its dissolution and precipitation patterns. We performed a 3D reconstruction of one of its nearly vertical walls by photogrammetry (see Methods~\ref{method:field}). Fig.~\hyperref[SaintMarcel]{\ref{SaintMarcel}A} shows the corresponding orthophotograph, on which spectacular 'scallops' of similar sizes and orientations can be observed. For a fairly homogeneous region of about $\SI{5}{\square\meter}$, we plot the wall elevation field $\eta(x,y)$ in Fig.~\hyperref[SaintMarcel]{\ref{SaintMarcel}B}. It consists of an array of concavities, a few dozen centimeters wide and a few centimeters deep, surrounded by narrower crests. Crests delimiting the scallops appear very well as areas of intense negative curvature forming thin contour lines on the corresponding mean curvature field $\kappa(x,y)$ (see Methods~\ref{method:curv} and Fig.~\hyperref[SaintMarcel]{\ref{SaintMarcel}C}). By plotting a longitudinal cross-section profile in Fig.~\hyperref[SaintMarcel]{\ref{SaintMarcel}D}, we visualize indeed that the curvature minima are precisely located at the top crests of the signal.\\

Schematically, these crests can be seen as singularities of the surface, \textit{i.e.} local discontinuities of the first derivatives $\partial \eta  / \partial x$ or $\partial \eta  / \partial y$, or equivalently divergences of the second derivatives and of the curvature. \\
In reality, in a physical case like this, singularities are regularized: divergences of the curvature are replaced by localized peaks. But although the radius of curvature at the crests is not zero, it is much smaller than the typical dimensions of scallops. On a large scale, it remains therefore relevant to describe the cave surface as a set of singular structures. As these are randomly distributed in space, we use two statistical methods to detect and quantify them, by computing the curvature distribution of the surface and its Fourier spectrum.\\

As observed in Fig.~\hyperref[SaintMarcel]{\ref{SaintMarcel}E}, the mean curvature distribution of $\eta(x,y)$ differs significantly from the centered Gaussian distribution of same standard deviation (pink solid line) in two respects: it has a tail at negative curvature and its maximum is shifted toward positive values. The tail shows the presence of very localized areas of strongly negative curvature: the crests. The maximum is shifted because most of the points are within a wide concave area of small but positive curvature. The probability density function is thus asymmetric, which can be evidenced by the computation of the skewness $ \tilde{\mu_3}= \left[\mathbb{E} \left(  (\kappa-\mu) /\sigma \right) \right]^3$, with $\mu$ and $\sigma$ the mean and the standard deviation of $\kappa$, respectively. We indeed find a negative skewness $ \tilde{\mu_3}=-0.33$.\\

As observed in Fig.~\hyperref[SaintMarcel]{\ref{SaintMarcel}F}, the power spectral density of the surface $S_\eta$ (see Methods~\ref{method:psd}) follows a clear power law in $k^{-4}$ at intermediate to low wavelength $\lambda=2\pi/k$. The existence of this power law can be linked to the presence of crests using an approach proposed initially to address the turbulent spectrum of steep waves~\cite{Kuznetsov2004}. Mathematically, singularities are scale-invariant objects which are known to have a very wide spectral signature, and more specifically a power law spectrum. The corresponding exponent can be intuited as follows. Let's assume that the interface is composed of randomly distributed spikes, whose first derivatives are discontinuous, surrounded by smoother zones. The Laplacian of the surface can then be approximated as the sum of randomly distributed Dirac delta functions, located at the position $\mathbf{r_i}$ in a 2D space, and of a regular function $f(\mathbf{r})$:
\begin{equation}
\dfrac{\partial^2 \eta}{\partial x^2} + \dfrac{\partial^2 \eta}{\partial y^2} = \sum_i\,\Gamma_i\,\delta (\mathbf{r}- \mathbf{r_i}) + f(\mathbf{r})  \, . 
\end{equation}\\
Assuming that the irregular part dominates the spectrum, by applying a 2D Fourier transform of wavevector $\mathbf{k}$ we find:
\begin{equation}
k^2\,\tilde{\eta} \sim \sum_i\,\Gamma_i \, \mathrm{e}^{-\imath\,(k_x\,x+k_y\,y)} \quad \quad \mathrm{with} \quad k^2=k_x^2+k_y^2 \, . 
 \end{equation}
The 2D Fourier transform of the surface therefore verifies $ \tilde{\eta} \sim k^{-2}$. According to the definition of the power spectrum $S_\eta(k)$ (see Methods), which is integrated over the directions, we obtain finally $S_\eta(k)=2\,\pi\,k\,|\tilde{\eta}|^2 \sim k^{-3}$. This power-law is thus associated to the presence of point-like singularities. Now if the spikes are better depicted by line singularities, the space spectrum becomes proportional to $k^{-4}$. By extension, if the crests have a non-integer 
fractal dimension $D$, we have $S_\eta \propto k^{-3-D}$~\cite{NazarenkoJFM2010}. Here, the observed power law is therefore compatible with the presence of one-dimensional crests. 

We have thus identified two indicators of the presence of crests on a natural wall. Although regularized on a small scale, these crests can be assimilated to singular structures, characterized by an asymmetric mean curvature distribution and a characteristic power-law in $k^{-4}$ in their Fourier spectrum. Because of the time scales required to form scallops on limestone (typically several thousand years), it is impossible to observe directly the evolution of the pattern and the appearance of singularities. However, in the following sections, we will present a numerical model and an experimental study to shed light on the generic emergence of scallops in erosion by dissolution. \\

\section{Normal ablation model}

\subsection{Theoretical basis of erosion by dissolution}

\subsection{Theoretical basis of erosion by dissolution}

Dissolution is a mass transfer phenomenon between a solid and a liquid phase, in which solid is transported in the liquid as a solute concentration field $c(x,y,z)$. Dissolution dynamics and solid morphogenesis thus depend, via the boundary condition at the interface, on solute transport by the flow. While the latter can be described by the classic advection-diffusion and Navier-Stokes equations, the macroscopic description of dissolution at the solid/liquid interface is slightly more complex. In the simplest modeling of chemical kinetics, the reaction rate can be shown to be proportional to the distance to the thermodynamic equilibrium~\cite{Colombani2007}(\textit{i.e.} $c=c_{\mathrm{sat}}$) leading to the following expression for the interface velocity $\mathbf{v_d}$:
\begin{equation}
- \rho_\mathrm{s}\,\mathbf{v_d}= \alpha (c_{\mathrm{sat}}-c_\mathrm{i})\,\mathbf{n}    
\label{cvd4}
\end{equation}
where $c_\mathrm{i}$ is the solute {mass} concentration at the fluid-solid interface, $\rho_\mathrm{s}$ the density of the solid, $\alpha$ a coefficient {with the same dimension as a speed} depending on the chemical properties of the involved solid/liquid system and $\mathbf{n}$ the unit vector normal to the interface. 

Then, conservation of the solute flux at the interface also gives~\cite{Cohen2020,Sharma2022,Chaigne2023}:
\begin{equation}
\rho_\mathrm{s}\,\mathbf{v_d}\,\left(1-\dfrac{c_\mathrm{i}}{\rho_\mathrm{i}}\right)=D\, (\mathbf{\nabla}c{\vert}_{\mathrm{i}} \cdot \mathbf{n} ) \,\mathbf{n} \, ,  \label{cvd5}
\end{equation}
with $\rho_\mathrm{i}$ the density of the fluid at the interface and $D$ the diffusion coefficient of the solute in the fluid.

In most cases for usual soluble materials (limestone, gypsum, salt, sugar ...), $\mathbf{v_d}$ is several orders of magnitude smaller than the typical velocity of the fluid, so that the interface can be considered as quasi-static. The hydrodynamics adapts to the new boundary conditions in a time far faster than those needed to have significant shape changes. In addition, if the chemical dissolution kinetics is fast compared to diffusion, $c_\mathrm{i}$  increases quickly to approach $c_{\mathrm{sat}}$~\cite{Philippi2019} and the concentration profile decreases to reach the value of the surrounding water bath on a small scale $\delta$, the thickness of the concentration boundary layer. The erosion rate -and potential patterning due to its heterogeneity- are controlled by the concentration gradient {(see Eq.~[\ref{cvd5}])} and thus by $\delta$, which results from the balance between diffusion and fluid advection and can be modulated by a coupling between the topography and the flow. To obtain the exact shape evolution of a dissolving solid in a specific case, the hydrodynamic problem should therefore be solved, either analytically or numerically, to access $\delta$. It often proves to be particularly difficult.\\
But the main point of this paper is to argue that the appearance of crests and spikes is largely independent of the exact behaviour of $\delta$. We show for instance that they can emerge even when $\delta$ is uniform, as long as the erosion velocity is normal to the solid interface, {which is the case according to Eq.~[\ref{cvd4}] and Eq.~[\ref{cvd5}]}.\\

\begin{figure}
   \centering
   \includegraphics[width=1\columnwidth]{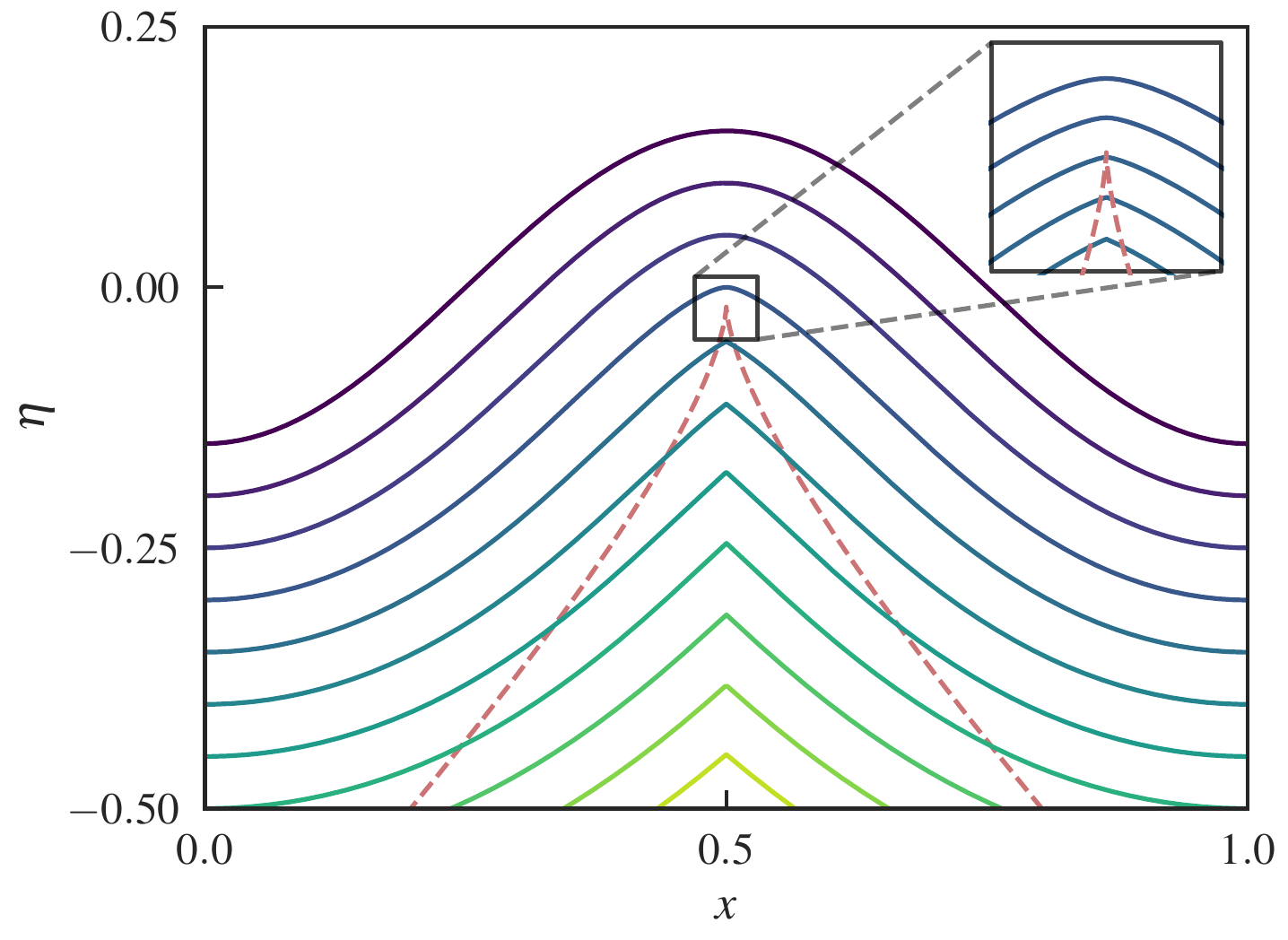}
   \caption{\label{Model0} Numerical model: normal ablation of a sinusoid with uniform dissolution rate. From top to bottom, successive interfaces resulting from the normal ablation, for selected steps. Pink dotted line: evolute curve of the initial sinusoid shape, \textit{i.e.} locus of its centers of curvature.}
\end{figure}

Let us first consider the simple case where the initial interface is one-dimensional, and more precisely a sinusoid. We compute the erosion in propagating each point of the initial interface by a fixed normal distance $d=0.05$ while ensuring that the new interface does not self-intersect (see SI Section 1. Algorithm of interface propagation). We repeat the process 10 times and plot the resulting interface at each step (see Fig.~\ref{Model0}). In the vocabulary of differential geometry, the successive interfaces drawn in this figure are said to be parallels of the initial interface. Parallel curves or also called offset curves have been the object of several studies, mainly in the context of computer-aided design~\cite{pham1992offset,maekawa1999overview}. A parallel curve does not usually have an analytical expression, even if this is the case for the initial curve. Yet if one considers a regular initial curve, parameterized by a parameter $s$, of curvature $\kappa_0(s)$, the curvature $\kappa_{d_{\mathrm{er}}}(s)$ of one of its parallel curves is simply expressed as:
\begin{equation}
    \kappa_{d_{\mathrm{er}}}(s)=\frac{\kappa_0(s)}{1+\kappa_0(s) \, d_{\mathrm{er}}},
\end{equation}
with $d_{\mathrm{er}}$ the distance between both curves \cite{farouki1990analytic}. Here, $d_{\mathrm{er}}=n_{\mathrm{it}} d$ with $n_{\mathrm{it}}$ the number of iterations. If $\kappa_0(s)>0$, then $\kappa_{d_{\mathrm{er}}}(s)$ decreases when $d_{\mathrm{er}}$ increases and tends toward 0: the convex portions of the sinusoid, located around the minima, become increasingly flattened. But if $\kappa_0(s)<0$, $\kappa_{d_{\mathrm{er}}}(s)$ decreases (its absolute value increases) and eventually diverges when $d_{\mathrm{er}}=-1/\kappa_0(s)$,
which means that $d_{\mathrm{er}}$ is equal to the radius of curvature. We plot the evolute of the initial sinusoid, which is the locus of its centers of curvature, on Fig.~\ref{Model0}. We indeed observe that a tip singularity appears when the parallel curve crosses the evolute, and the location of the tip coincides with the intersection point.
Alternatively, a singularity of $\eta(d_{\mathrm{er}},x)$ can be seen as a shock of $\partial_x \eta$. Indeed, in the limit of small deformations, $\partial_x \eta$ follows the inviscid Burgers equation, which is well known to give rise to shocks, \textit{i.e.} discontinuities of $\partial_x \eta$; and $\eta$ follows the deterministic Kardar-Parisi-Zhang equation without surface tension~\cite{KardarPRL1986}.
{Shock formation has also been theoretically investigated in the context of crystal growth~\cite{Tsemekhman2002,Tsemekhman2003} and ice crystal melting~\cite{cahoon2006growth}.}
With the exception of flat and spherical shapes, most interfaces whose curvature changes sign and subjected too a normal ablation process will form in finite time sharp crests corresponding to singular points where the surface gradient is discontinuous. We emphasize thus that generally a random surface experiencing normal ablation will exhibit the same singularities, as shown in the next section. 

\subsection{Uniform dissolution rate for a 1D interface}

\begin{figure*}[t!]
   \centering
   \includegraphics[width=0.88\textwidth]{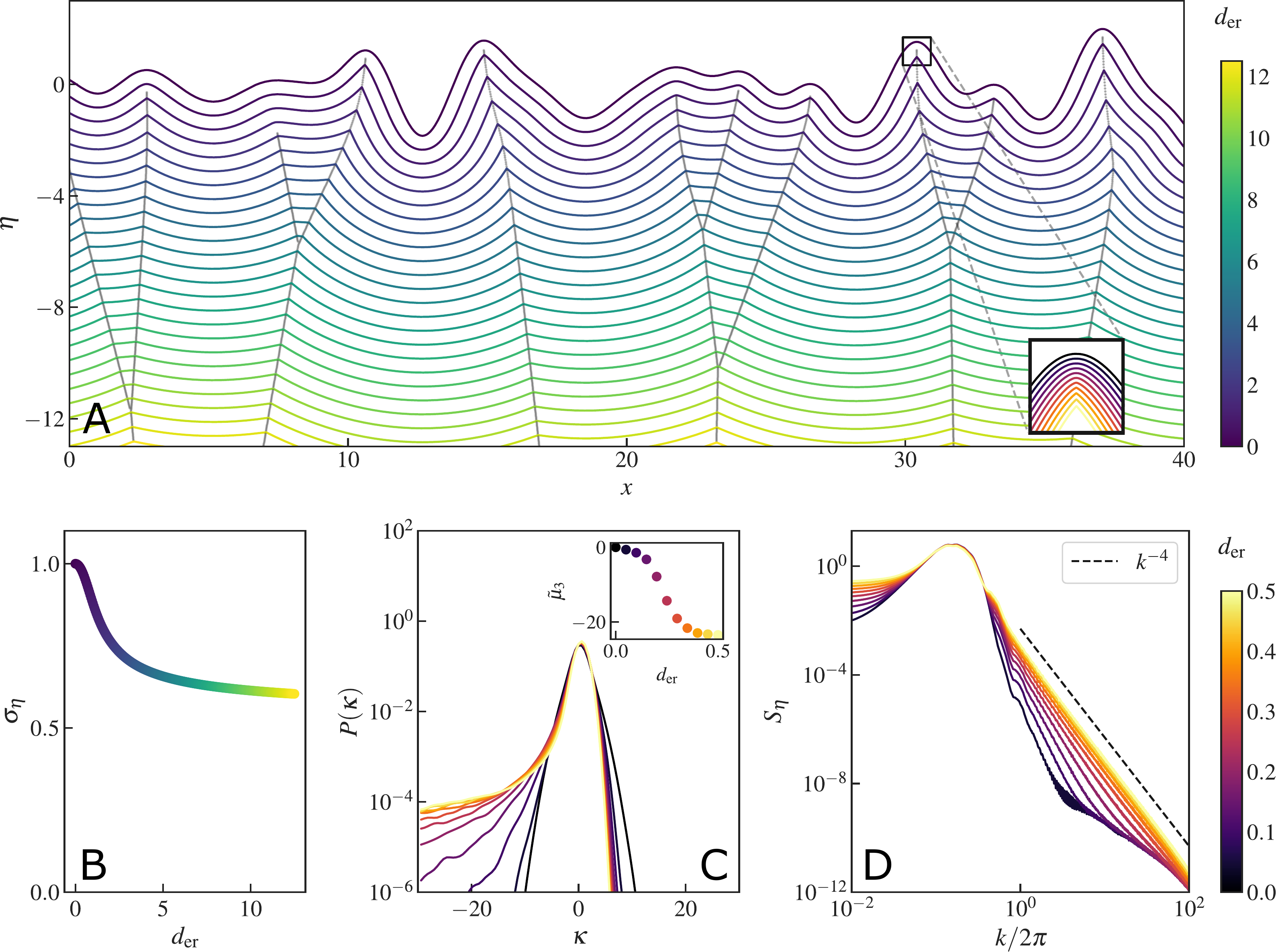}
   \caption{\label{Model1} Numerical model: normal ablation of a random interface with a uniform dissolution rate. A: Successive interfaces resulting from the uniform ablation of a random initial interface, consisting of 10$^6$ points in the interval $[-500,500]$. The location of tips is highlighted with grey dots. Inset: focus on the first interfaces with emergence of a tip. B: The standard deviation of the interface $\sigma_{\eta}$ decreases with erosion distance $d_{\mathrm{er}}$. C: The curvature distribution becomes asymmetric with a tail at highly negative $\kappa$ evidencing the apparition of tips. We focus on the beginning of the process, and the color code is the same as in the inset of panel A. Inset: skewness of the distribution $\tilde{\mu_3}$ as a function of $d_{\mathrm{er}}$. Note that in absence of regularizing mechanisms of singularities, $\tilde{\mu_3}$ reaches here strongly negative values.  D: The power spectral density $S_{\eta}$ of the interface rapidly collapses at small wavelengths on a characteristic power-law in $k^{-4}$.}
\end{figure*}

Let us now consider what happens when a random interface undergoes uniform normal ablation in the one-dimensional case. We model the interface by a set of points $\left(x,\eta(x)\right)$ forming a line. We generate the initial interface by applying Gaussian filters on a random list to select dimensionless wavelengths between 1 and 1.5, and we set its standard deviation to 1. Then, we obtain the subsequent interface by propagating each point in the direction of its normal by a given length $d=0.05$. We repeat the process 250 times and plot the interface once every 10 iterations in Fig.~\hyperref[Model1]{\ref{Model1}A}. The color indicates the total ablation length $d_{\mathrm{er}}=n_{\mathrm{it}} d$, with $n_{\mathrm{it}}$ the number of iterations.\\
Because normal vectors deviate from each other at the local minima, we see that the troughs widen and their curvature decreases. At the same time, as the normal vectors converge at the local maxima, humps narrow with each iteration until they form tips (see inset of Fig.~\hyperref[Model1]{\ref{Model1}A}). Adjacent tips eventually merge which leads to a global coarsening of the pattern, as evidenced by the grey dots indicating the position of tips in Fig.~\hyperref[Model1]{\ref{Model1}A}.
Finally, the amplitude of the height variations of the interface decreases. We highlight this by plotting $\sigma_{\eta}$ the standard deviation of $\eta$ as a function of $d_{\mathrm{er}}$ in Fig.~\hyperref[Model1]{\ref{Model1}B}.\\
Then, focusing more precisely on the beginning of the process, we plot the curvature distribution of the first ten interfaces in Fig.~\hyperref[Model1]{\ref{Model1}C}, and notice that it becomes increasingly asymmetric. The apparition of a tail on the left part of the distribution shows the emergence of points with a highly negative curvature: the tips. The skewness of the distribution, plotted versus the ablation length in inset, decreases sharply for $d_{\mathrm{er}}\simeq 0.25$. The inflection point seems to coincide with the apparition of the first singularities. Concurrently, we plot the power spectral density $S_{\eta}$ as a function of the inverse of the wavelength in Fig.~\hyperref[Model1]{\ref{Model1}D}. The contribution to the signal of small wavelengths increases gradually, and the right part of the spectrum eventually converges towards a power-law in $k^{-4}$, characteristic of singular structures (see Section 1). The exponent is expected to be $-4$ because, unlike crests on a surface, tips are point-like singularities, but there is no integration over the directions in 1D.\\

{This model explains the evolution of any corrugations towards sharp tips, in line with the results of Kardar, Parisi and Zhang. Indeed, the KPZ equation predicts that for a normal growth process, a one dimensional interface evolves into a juxtaposition of paraboloid segments separated by discontinuities of the gradient, the shocks~\cite{KardarPRL1986}. These paraboloids then progressively merge which leads to coarsening. These results also apply for the reverse process, ablation. However, neither our model nor the KPZ equation can explain the very emergence of a pattern: they both predict a decrease of the corrugation amplitude.} Pattern formation necessitates differential dissolution. In the following subsection, we will therefore introduce a non-uniform dissolution rate on a two-dimensional interface to allow three dimensional patterning as observed in natural systems.

\begin{figure*}[t]
   \centering
   \includegraphics[width=0.98\textwidth]{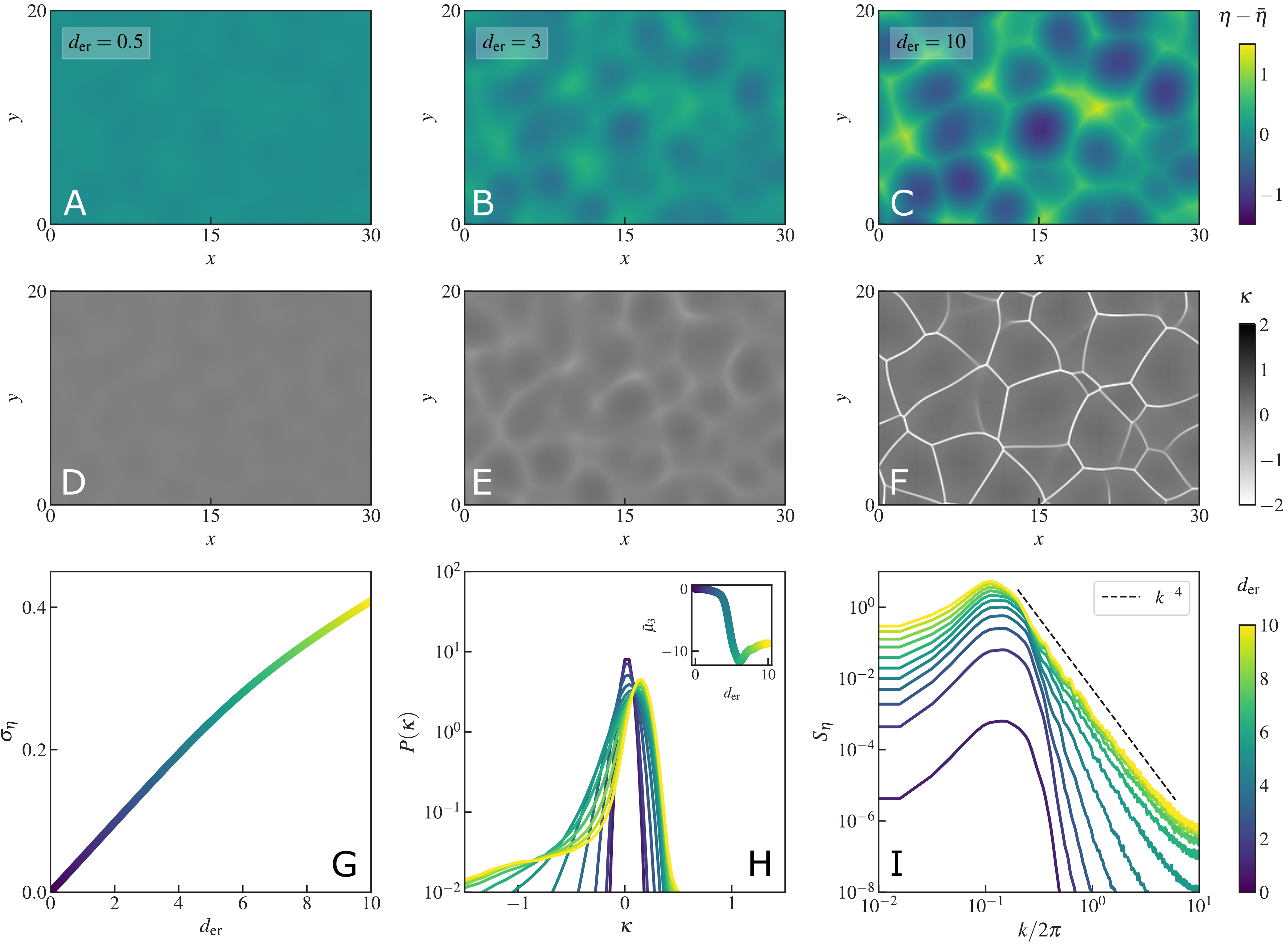}
   \caption{\label{Model2} Numerical model: normal ablation of an initially flat surface with non-uniform dissolution rate. A-C:  Topography of the surface at three different iterations (only a detail is shown and the average value of $\eta$ is subtracted for clarity). D-F: Corresponding mean curvature field. G: Standard deviation of the topography versus erosion length. H: Distribution of the mean curvature. It becomes increasingly asymmetric as $d_{\mathrm{er}}$ increases. Inset: skewness of the distribution. I: The power spectral density of the topography collapses at small wavelength on a characteristic power-law in $k^{-4}$. }
\end{figure*}

\subsection{Heterogeneous dissolution rate for a 2D interface}

We start from a flat surface $\eta(x,y)=0$ consisting of a square grid of $1800\times1800$ points. We obtain the subsequent interface by propagating each point in the direction of its normal by a now spatially-varying length $d(x,y)$, and we then repeat the process. We introduce space correlation so that, at each iteration,
\begin{equation}
    d(x,y)=d\left(1+\xi(x,y)\right),
\end{equation}
with $d=0.05$ (very small compared to the other length scales) and $\xi(x,y)$ a random function, varying between $-0.4$ and $0.4$, obtained by applying Gaussian filters on a random matrix to select wavelengths between 2 and 3. The spatial variations in $d(x,y)$ first imprint on the interface and create a smoothly varying pattern (see Fig.~\hyperref[Model2]{\ref{Model2}A}). Then, because the ablation is normal, troughs become wider and humps become narrower (Fig.~\hyperref[Model2]{\ref{Model2}B}) until they form lines of crests. At later times, we recover the main features of scallop patterns: sharp crests encircling broad concave zones, forming a cellular pattern at large scale (Fig.~\hyperref[Model2]{\ref{Model2}C}). On Fig.~\hyperref[Model2]{\ref{Model2}D-F}, we plot the mean curvature at each point of the surfaces shown above. A network of crests appears progressively, characterized by a very negative mean curvature.\\
While the amplitude of the pattern continuously increases with time (Fig.~\hyperref[Model2]{\ref{Model2}G}), the two indicators of singularities identified in the previous sections hold. The curvature distribution becomes increasingly asymmetric with a tail at highly negative mean curvature and a maximum shifted towards positive values. Its skewness decreases first slowly then abruptly, before reaching a plateau (Fig.~\hyperref[Model2]{\ref{Model2}H}). At small wavelength, the power spectral density scales like $k^{-4}$ evidencing one-dimensional singular structures (Fig.~\hyperref[Model2]{\ref{Model2}I}).\\

We were thus able, with a simple model containing no hydrodynamics, to reproduce scallop patterns. All that is required is to impose a normal ablation and a heterogeneous ablation rate correlated in space. Then, even though the ablation rate varies smoothly, a cellular pattern with sharp crests robustly appears. We therefore emphasize that the tip formation mechanism mentioned in the previous subsections is robust and holds when erosion is no longer uniform. The average size of the cells, which initially reflects the typical wavelength of the ablation rate variations, increases as cells merge; it results in a coarsening of the pattern. It can be noted that the same results are obtained, qualitatively, if $\xi(x,y)$ varies at each iteration instead of being constant. But the shorter the correlation time, the slower the amplitude growth. The whole hydrodynamics is in fact hidden behind two parameters: the typical wavelength of the erosion rate variations and the typical "correlation time". In a physical case, they would be typically imposed by the hydrodynamic instability involved and by the specific feedback between the flow and the topography.\\
In the following section, we demonstrate the relevance of this model to explain the appearance of scallops with an experimental example. A hydrodynamic instability developing at a water-soluble material interface produces a spatially varying dissolution rate. In accordance with the model, scallops appear consequently on the material surface.

\begin{figure*}[t]
   \centering
   \includegraphics[width=0.96\textwidth]{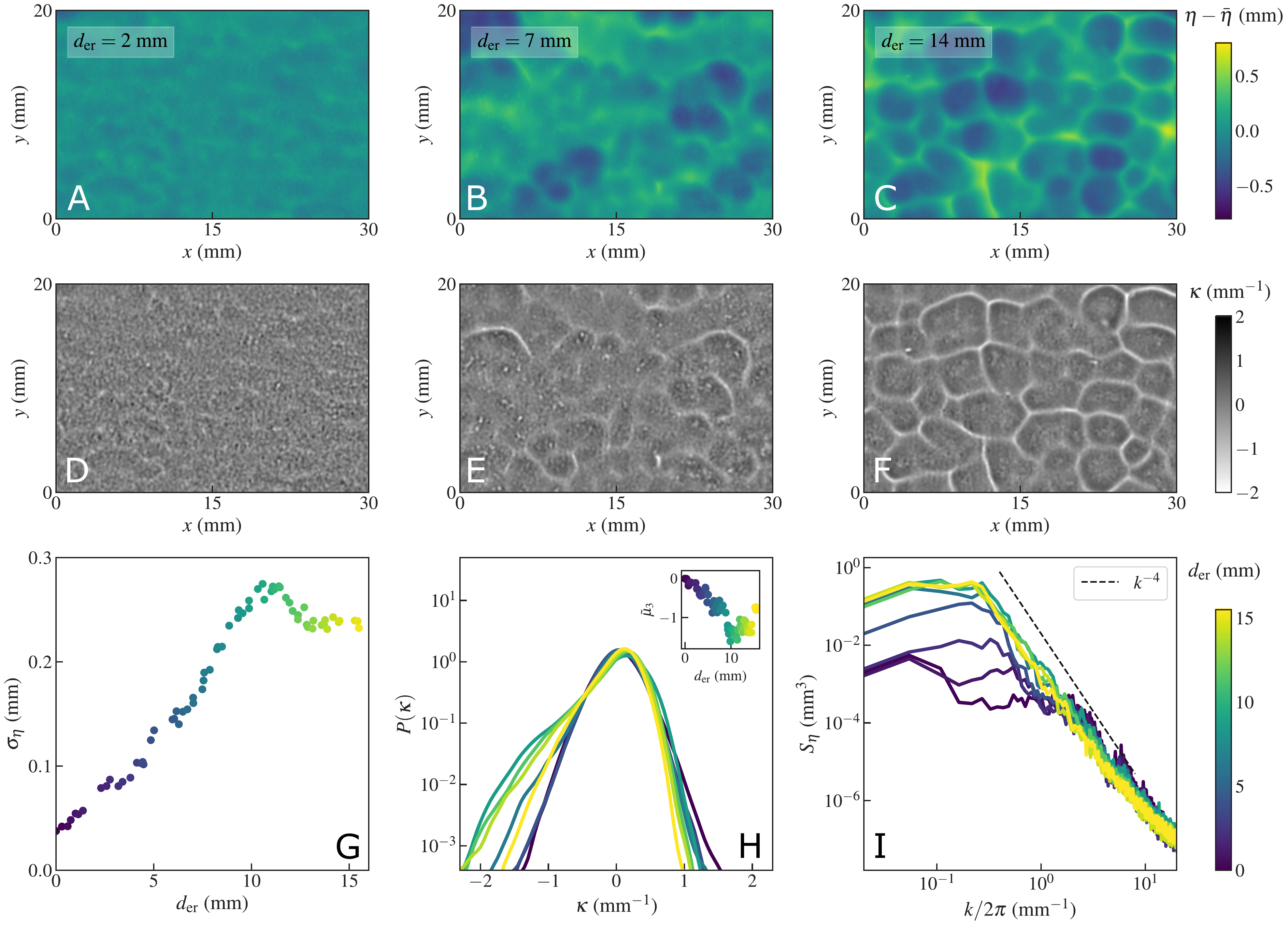}
   \caption{\label{ExSaumure} Experiment: dissolution of the bottom surface of a salt block in a water tank. A-C: Topography of the surface at three different moments (only a detail from the center of the block is shown and the average value of $\eta$ is subtracted for clarity). D-F: Corresponding mean curvature field. G: Standard deviation of the topography versus erosion length. H: Distribution of the mean curvature. It becomes increasingly asymmetric as $d_{\mathrm{er}}$ increases. Inset: skewness of the distribution. I: The power spectral density of the topography collapses at small wavelength on a characteristic power-law in $k^{-4}$. }
\end{figure*}

\section{Comparison to scallop patterns in solutal convection experiments.}

In order to experimentally study the emergence of scallops in a simple and controlled way, we suspend horizontally blocks of soluble material in a large aquarium filled with water. The blocks of Himalayan pink salt of dimensions $20\times10\times\SI{2.5}{\cubic\centi\meter}$ are polished with abrasive papers of decreasing roughness so that their bottom surface is initially flat and smooth. Once placed in water, blocks start to dissolve and a salt-rich concentration layer grows around the solid. Below the block and once its thickness reaches a critical threshold $\delta_{\mathrm{c}}$, this dense layer is subjected to a solutal Rayleigh-Bénard instability~\cite{Philippi2019}. It then destabilizes into plumes sinking in the bottom of the tank. The characteristic distance between plumes scales like $\delta_{\mathrm{c}}$, which is given by a constant Rayleigh number criterion~\cite{Sullivan1996,Philippi2019,Cohen2020,Sharma2022}:
\begin{equation}
\frac{\Delta\rho~g~\delta_{\mathrm{c}}^3}{\eta D}=\mathrm{Ra_c},
\label{eq:rayleigh}
\end{equation}
with $\Delta \rho$ the difference of density between the saturated fluid and the fluid in the bath, $\eta$ the dynamic viscosity of the fluid, $D$ the diffusion coefficient of the solute in the fluid, and $\mathrm{Ra_c}$ a critical value of the solutal Rayleigh number. \\

Now, because the thickness of the concentration boundary layer is locally larger at the vertical of a plume than between two plumes (where fresh water comes up), the dissolution rate is heterogeneous according to Eq.~[\ref{cvd5}], which could induce a patterning of the solid surface\cite{Cohen2020}. We therefore regularly take the blocks out of the tank and scan their lower surface with a laser profilometer, in order to access their topography with an accuracy of a tenth of a millimeter.\\

Let us now present the results obtained in a typical experiment. The aquarium was filled with an almost-saturated brine ($\rho_0=\SI{1.195}{\kilo\gram\per\liter}$), in order to increase the size of the critical boundary layer thickness (see Eq.~[\ref{eq:rayleigh}]) and thus the size of the patterns\cite{Cohen2020}. The bottom surface of the salt block first becomes increasingly rough. After an average erosion depth of $\SI{2}{\milli\meter}$, the surface indeed displays height variations of typically $\SI{0.1}{\milli\meter}$. It can be seen on Fig.~\hyperref[ExSaumure]{\ref{ExSaumure}A}, where the topography of a $20\times\SI{30}{\square\milli\meter}$ area of the surface is shown. Then, cavities appear and broaden (see Fig.~\hyperref[ExSaumure]{\ref{ExSaumure}B}). After an average erosion depth of $\SI{14}{\milli\meter}$, the surface is covered by scallops, of a few millimeters width, forming a cellular pattern (see Fig.~\hyperref[ExSaumure]{\ref{ExSaumure}C}). As can be seen on the maps showing the mean curvature (see Fig.~\hyperref[ExSaumure]{\ref{ExSaumure}D-F}), the cavities are surrounded by increasingly narrow crests that progressively merge into a connected network.\\

The amplitude of the pattern first increases, almost linearly. Yet as shown Fig.~\hyperref[ExSaumure]{\ref{ExSaumure}G}, it eventually passes through a maximum before decreasing slightly to reach a plateau. In order to explain the initial phase, we assume that the crests channel the plumes so that their position is locked. We are thus in the case of the model proposed previously, where the erosion rate is spatially variable (maximum at the level of the troughs and minimum at the level of the crests from where the salt-laden plumes flow) but constant in time. {A possible explanation to the fact that the pattern then stops growing could be the competition between differential dissolution, which initiates the pattern, and normal ablation, which reduces the amplitude once crests are formed (see Fig.~\hyperref[Model1]{\ref{Model1}B}). However, a saturation of the pattern amplitude is not observed in the model with heterogeneous dissolution rate (see Fig.~\hyperref[Model2]{\ref{Model2}G}): there is only a slowdown in growth. Therefore, we believe that this competition cannot entirely explain saturation. In the experiment, there is a feedback mechanism between the topography and the flow that modifies the spatial variations in boundary layer thickness, and hence in dissolution rate, once patterns are formed. Such a feedback is not taken into account in our numerical model, but should probably be considered in order to explain and predict the final amplitude of the pattern. The physical mechanism involved will be the subject of a future study.
}\\

What is of particular interest to us here is that we once again find the two indicators of sharp structures. The mean curvature distribution of the surface becomes more and more asymmetric as dissolution proceeds (see Fig.~\hyperref[ExSaumure]{\ref{ExSaumure}H}). Its skewness, initially zero, decreases significantly before reaching a plateau. Finally, the power spectral density converges, at small wavelength, to the characteristic power-law in $k^{-4}$ (Fig.~\hyperref[ExSaumure]{\ref{ExSaumure}I}).   

\section{Discussion and conclusion}

In this article, we demonstrate that the common generation of sharp shapes appearing in nature by dissolution erosion corresponds to the emergence of tip singularities in the evolution of the interface. We first show that a statistical analysis of natural dissolution patterns reveals the presence of tip singularities. We then explain their appearance by a simple model of normal ablation. First considering the basic case of a one-dimensional interface and uniform erosion rate, we show that initial humps thin until forming tips, while troughs widen. The first singularities appear when the moving interface crosses the evolute. Therefore, any surface presenting curvature variations will evolve to display singularities in finite time. 
This approach follows the idea of Arthur Lange, who was the first to propose a one-dimensional process of uniform ablation to explain the sharp shapes observed in caves~\cite{Lange1959}. This mechanism was then evoked qualitativelly in the formation of dissolution flutes and scallops~\cite{Curl1966}. {The emergence of singularities as discontinuities of the interface gradient is also very similar to shocks appearing in interfaces whose evolution is governed by the deterministic Kardar-Parisi-Zhang equation~\cite{KardarPRL1986}}.

\begin{figure}[t!]
   \centering
   \includegraphics[width=0.93\columnwidth]{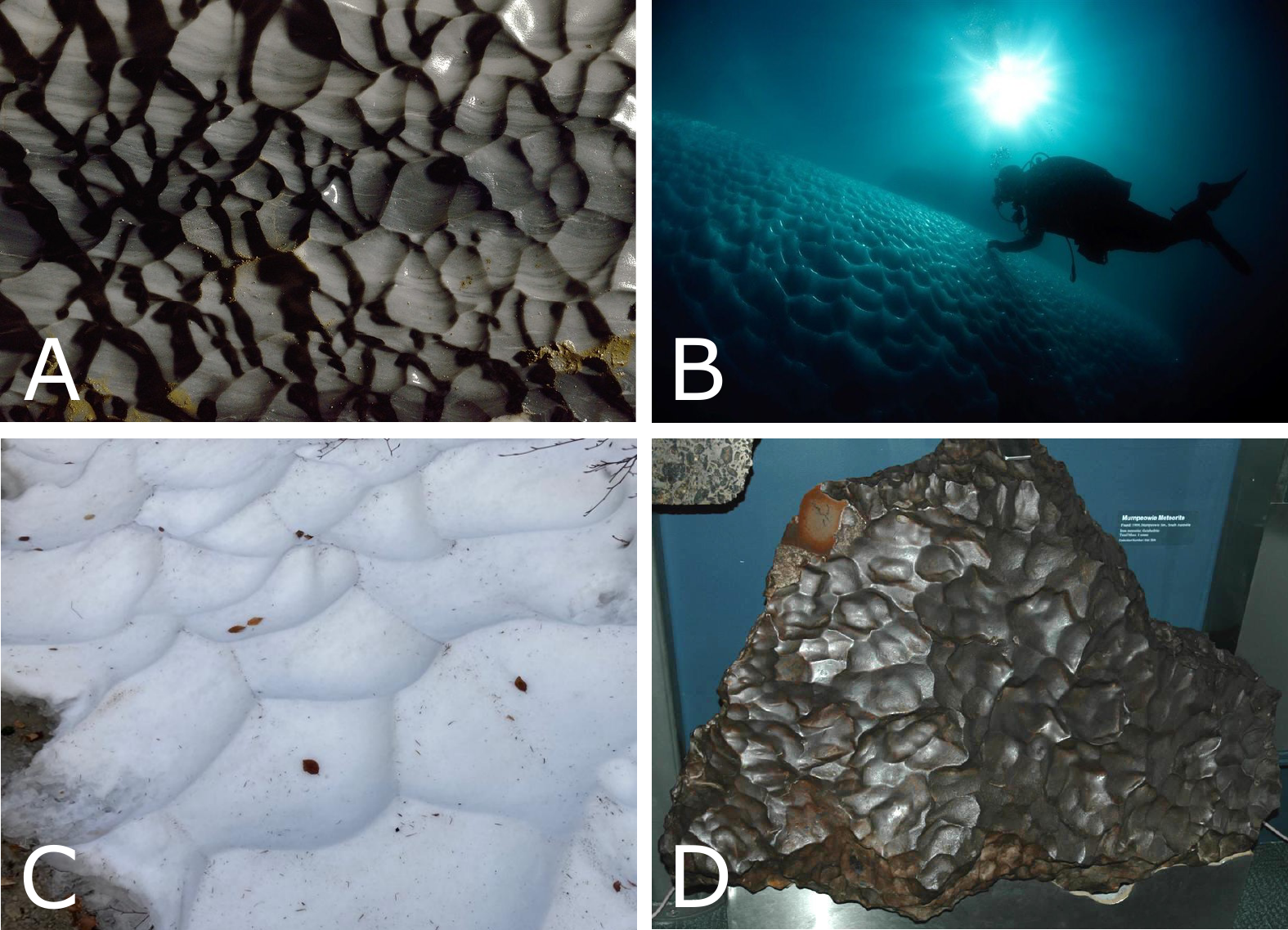}
   \caption{\label{SpikesExamples} Natural examples of spike and scallop patterns. A: Sharp scallops on marble limestone from the Korallgrottan cave in Sweden. Width about 0.5 m. Credits Johannes Lundberg. B: Ice scallops on the immersed surface of an iceberg. Credits Le cinquième rêve - from the film "Le piège blanc" directed by Thierry Robert. Underwater images by René Heuzey. C: Suncups formed by sublimation of snow in Vercors, France.  Credits \'Emilien Dilly. D: Scallop patterns on the Murnpeowie Meteorite, South Australian Museum, Adelaide, Australia. Width about 1 m. Credits James St. John.}
\end{figure}

Yet a uniform ablation model cannot explain the emergence of a pattern. We therefore propose a new model, this time two-dimensional, in which the erosion rate varies spatially with a given typical wavelength. We first logically observe that these variations in the erosion rate are imprinted on the surface, thus creating a pattern. Then, hills become narrower and troughs become wider, until a cellular pattern of cavities surrounded by sharp crests is obtained. This pattern is strikingly reminiscent of the scallops observed in limestone caves and described in the first section (see for another example Fig.~\hyperref[SpikesExamples]{\ref{SpikesExamples}A}). Finally, we set up a simple experimental device to obtain an erosion rate varying spatially with a well defined wavelength. To do this, we immerse a block of salt in an aquarium and look at the evolution of its bottom surface. Underneath it, a solutal convection instability leads to the emission of plumes which induce a variable erosion rate. We then observe the appearance of scallops, quite similar to the natural ones, in accordance with the predictions of the previous model.\\
In all three cases, the statistical analysis of the interface reveals the presence of singular structures (even if the singularities are in fact regularized on a small scale). {These singularities are due to the fact that the erosion rate is at any point normal to the interface. Even though normal ablation does not necessarily imply the emergence of singularities (an ice cube melting in water will rather become rounder), we expect this to always be the case in the context of pattern formation. Indeed, in this case, the ablation rate must be minimal at the location of humps for the pattern to grow. If we add to this normal ablation, we have a sufficient condition for the formation of singularities: they will appear where the ablation rate is locally minimal.} In two dimensions, singularities get robustly organized in interconnected crest lines surrounding concave areas (\textit{i.e.} scallops). Therefore, every soluble surface subjected to a non-uniform erosion rate varying on a typical scale is expected to evolve into scallops. This typical scale could be selected, for instance, by any hydrodynamic instability. It probably explains why scallops appear so often, in a wide variety of materials and hydrodynamic conditions.\\

To conclude, we note that we have focused our study on the case of dissolution, which allows us to perform field measurements, experiments and numerical modeling, but our findings apply to a larger class of ablation phenomena. In particular, scallops patterns are also reported for melting interface and specifically for the ice/water interface in presence of a strong longitudinal current~\cite{gilpin_hirata_cheng_1980,Bushuk2019} or of gravity driven convection flows~\cite{Solari2013,Cohen2016,Favier2019,Weady2022} (see for example Fig.~\hyperref[SpikesExamples]{\ref{SpikesExamples}B}). Analog to the dissolution flutes~\cite{Curl1966,Claudin2017,Carpy2023}, the sublimation of icy substrates under turbulent flows generates ripples, which can be observed on Earth and other planets~\cite{bordiec2020sublimation,Carpy2023}. In different hydrodynamic conditions, cellular patterns created by sublimation are also reported like the sun cups on high altitude glaciers~\cite{betterton2001theory} and the ablation hollows on snow~\cite{jahn1968origin} (see Fig.~\hyperref[SpikesExamples]{\ref{SpikesExamples}C}). Mechanical abrasion may also produce similar shape like in the ablation at high temperature~\cite{vignoles2010modelling}. We note finally the striking resemblance of scallop patterns with the finger-like imprints at the surface of meteorites fallen on Earth, which are called Regmaglypts~\cite{Lin1986}, like the one depicted in Fig.~\hyperref[SpikesExamples]{\ref{SpikesExamples}D}. These patterns result from the ablation of meteorite surface during their entry in the atmosphere.

\section*{Methods}

\subsection{Field protocol}
\label{method:field}
The 3D reconstruction was done by photogrammetry. 118 pictures were acquired on the wall with a Nikon D750 camera. The light was provided by two spotlights placed on each side of the scene. The spotlights were intentionally not placed directly in front of the wall but with a slight angle. As the rock is very white, the studied landforms were almost invisible with direct light. Two different rulers were placed on the wall for scaling the model. The DTM (Digital Terrain Model) processing was done with Agisoft Metashape software.

\subsection{Surface characterization}

\subsubsection{Curvature field computation}
\label{method:curv}
If $\mathbf{n}$ is the local normal vector to the surface, the mean curvature is defined as $\kappa=- \nabla \cdot \mathbf{n}$. With our parametrization, the mean curvature is computed as:
\begin{equation}
\label{meancurvature}
\kappa=\dfrac{(1+\eta_y^2)\,\eta_{xx}+(1+\eta_x^2)\,\eta_{yy}-2\,\eta_x\,\eta_y\,\eta_{xy}}{(1+\eta_x^2+\eta_y^2)^{3/2}} \, ,
\end{equation}
with the notation $\eta_{x_i}$ and $\eta_{x_i x_j}$ for the first and second derivative relatively to the coordinates $x_i$ and $x_j$, respectively.
Nevertheless, for experimental signals the computation of first and second derivative are known to amplify the noise at small scale. We apply thus low-pass Gaussian filter to the field $\eta$  before.

\subsubsection{Space Power spectrum computation}
\label{method:psd}
After subtracting a second order fit, to remove the large scale structure and avoid biases due to an imperfect leveling, we compute the spatial-power spectrum of the the topography, $S_\eta$ (or power spectral density, PSD), after integrating along the different directions:
\begin{align}
\MoveEqLeft{ S_\eta(k_x,k_y)= \dfrac{1} {L_{x}L_{y}} \, \left\lvert \int_{0}^{L_{y}}\int_{0}^{L_{x}}\eta(x,y)\,\mathrm{e}^{-i(k_{x}x+k_{y}y)}\,\mathrm{d}x\,\mathrm{d}y\right\rvert^{2} }\\
\MoveEqLeft{  S_\eta(k)=\int_{0}^{2\pi}\,S_\eta(k_x,k_y)\,k \, \mathrm{d}\theta}  \label{Setaform} 
\end{align}
with $k_{x}=k\cos{\theta} , k_{y}=k\sin{\theta}$. In order to better compare with the physical dimensions, it can be useful to plot the spectrum as a function of $k/2\pi= 1/\lambda$.\\

\begin{acknowledgments}
We acknowledge for scientific discussions Etienne Couturier (MSC), Marc Durand (MSC), Olivier Devauchelle (IPG Paris). We thank for the field data acquisition in the Saint-Marcel cave, Olivier Bourgeois (LPG) and Delphine Dupuy (curator of the cave). This research was funded by the ANR grants Erodiss ANR-16-CE30-0005 and PhysErosion ANR-22-CE30-0017 as well as the Idex Emergence Grant Riverdiss (ANR-18-IDEX-0001) from Universit\'e Paris Cit\'e. The field measurements were funded by the Tellus program "Caractérisation et modélisation des formes de dissolution périodiques sur les parois calcaires soumises à des écoulements d'eau” project." from CNRS (INSU). This project was also supported by the Observatory of the sciences of the Universe Nantes Atlantique (Osuna, Nantes Université, CNRS UAR-3281, Université Gustave Eiffel, Conservatoire National des Arts et Métiers, Université d'Angers, Institut Mines-Télécom Atlantique)
\end{acknowledgments}


%

\newpage

\appendix

\section{Algorithm of interface propagation}

\begin{figure*}
\centering
\includegraphics[width=1\textwidth]{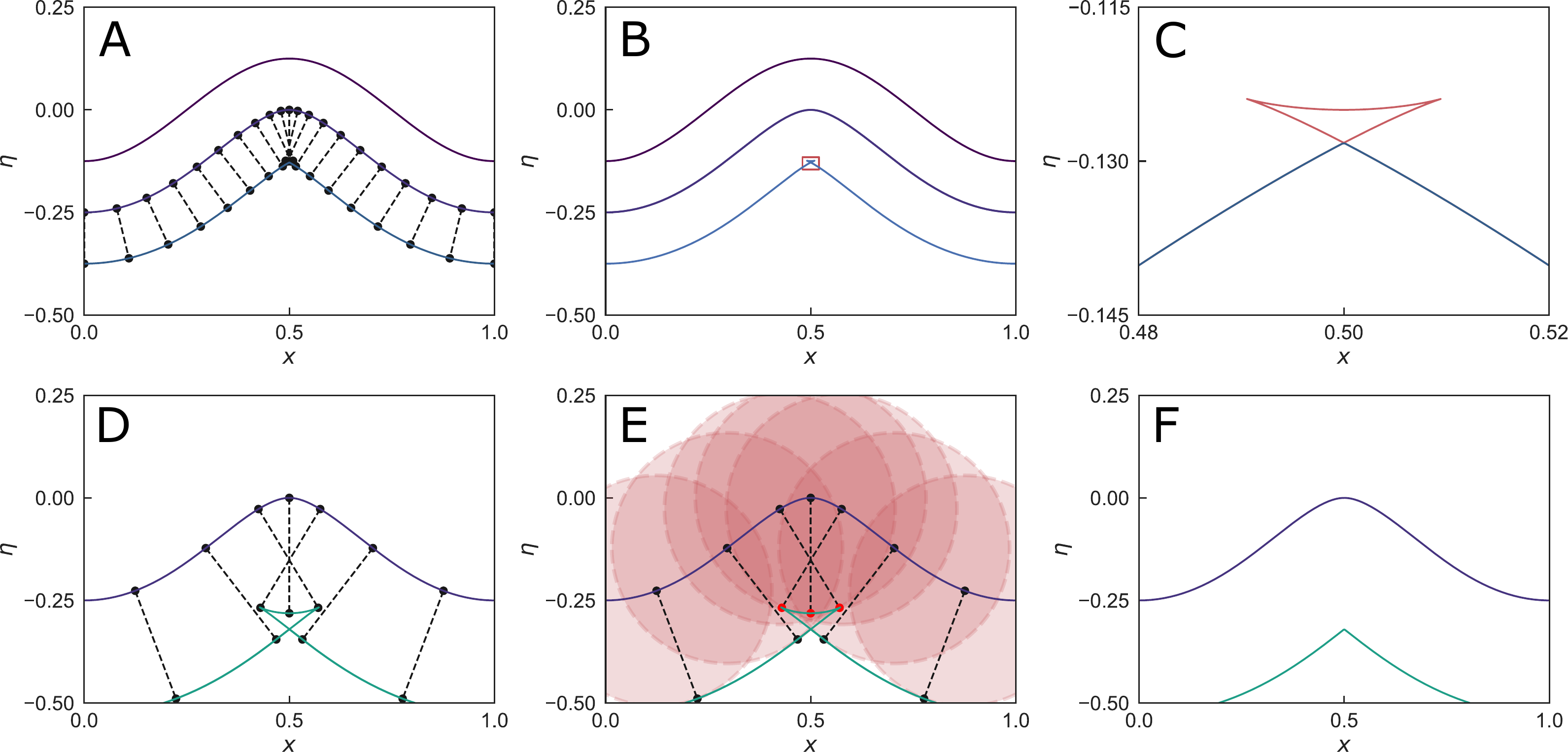}
\caption{Explanation of the principle of the algorithm, and how it removes loops or swallowtails:  A: Illustration of the principle of the interface propagation algorithm, in its naive implementation.  B: A loop appears at points where the radius of curvature is smaller than the erosion depth $d_{\mathrm{er}}$. C: Zoom in on the loop: the part in red should be removed from the new interface since it belongs to an area which has already been eroded. D: In green, curve obtained by the naive algorithm with a big loop. The points to remove are in fact points that are closer than $d_{\mathrm{er}}$ from the previous interface. E: The improved algorithm computes the distance between each point of the new interface and the points of the previous interface. If one of these distances is smaller than $d_{\mathrm{er}}$ (on the plot, if the point is inside one of the pink disks), then the point is removed from the new interface (here, coloured in red). F: In green, interface obtained with the improved algorithm, with no loop but with a tip.}
\label{fig:algo}
\end{figure*}

Applying a naive implementation of the algorithm leads to the formation of loops at points where the radius of curvature is smaller than the erosion depth $d_{\mathrm{er}}$, \textit{i.e} where the interface crosses the evolute. These loops are often referred to as "swallow-tails". Since the interface between two phases cannot self-intersect, they are not physically relevant and should be removed, a process which is called "trimming the offset curve" in the vocabulary of computer-aided design \cite{farouki1990analytic} or "applying Huygens' construction" \cite{sethian1996fast,sethian1999level}. To do so, we notice that the distance between points belonging to a loop and the initial curve is strictly inferior to $d_{\mathrm{er}}$, whereas it is equal to $d_{\mathrm{er}}$ for all other points. We therefore compute the distance between each point of the new interface and points of the initial interface, and remove the point if one of theses distances is strictly inferior to $d_{\mathrm{er}}$. The procedure is explained and illustrated on Fig.~\ref{fig:algo}. In reality, all distances between points of the new interface and points of the initial one are not computed: that would be uselessly time-consuming. Only distances between a given point of the initial interface and the neighbours of the point it comes from need to be computed. The number of neighbours that has to be taken into account depends on the ratio between the distance between two adjacent points and $d_{\mathrm{er}}$.  \\
For a two-dimensional interface, the principle of the algorithm is very similar. Swallow-tails still appear, and points belonging to a swallow-tail are removed \textit{a posteriori} by first computing the distance between points of the initial interface and points of the new interface, and then applying the same criterion as before. When the erosion rate is heterogeneous, \textit{i.e.} $d_{\mathrm{er}}=d_{\mathrm{er}}(x,y)$, the criterion that must be used to remove points belonging to a swallow-tail slightly changes. The distance between a point of the new interface of coordinates $(x',y',z')$ and all points of the initial interface of coordinates $(x,y,z)$ must always be superior to $d_{\mathrm{er}}(x,y)$ for the new point to be kept.  \\

\section{Experimental protocol}

The experiment is performed in a $\SI{40}{\centi\meter}\times\SI{20}{\centi\meter}$ aquarium filled with an almost-saturated brine ($\rho_0=\SI{1.195}{\kilo\gram\per\liter}$) to a height of $\SI{20}{\centi\meter}$. The blocks of dimensions $\SI{20}{\centi\meter}\times\SI{10}{\centi\meter}\times\SI{2.5}{\centi\meter}$ consist of Himalayan rock salt, extracted from a mine in Pakistan. They are first polished with abrasive papers of decreasing roughness so that their bottom surface is initially flat and smooth; and then placed in the aquarium. Every $\SI{3}{\hour}$, they are taken out of the brine and dried with compressed air. The topography of their bottom surface is measured by two complementary optical methods. First, the topography of the whole block is measured with a laser line scanner scanCONTROL 2900-100/BL from Micro-Epsilon3D\texttrademark, which gives access, by triangulation, to the profile $\eta(x)$ over a line; with a vertical accuracy of $\SI{70}{\micro\meter}$. The scanner is associated to a motorized translation stage LTS300 Thorlabs\texttrademark, which allows to repeat the measure periodically and to obtain the elevation $\eta(x,y)$ over the whole block. This data is used to compute the average erosion rate $d_{\mathrm{er}}$ and the amplitude $\sigma_\eta$ of the topography. Then, a $\SI{4.3}{\centi\meter}\times\SI{2.8}{\centi\meter}$ area located in the center of the block is scanned with LMI Gocator 3506\texttrademark, which gives access to $\eta(x,y)$ on a square grid of $\SI{20}{\micro\meter}$ pitch. The latter method is more precise and gives more details, but has a smaller field of view. This data is used to plot the the elevation and mean curvature on Fig~5A-F of the main text, and to compute the mean curvature distribution $P(\kappa)$ and power spectral density $S_\eta$ in Fig~5H-I.
A small amount of fresh water is added in the tank to compensate for the input of salt due to dissolution and ensure that the density of the brine remains constant. Finally, the block is placed back into the brine and the process is repeated. The block remains immersed for a total of about 225 hours before being almost completely dissolved.

\begin{figure*}
\centering
\includegraphics[width=1\textwidth]{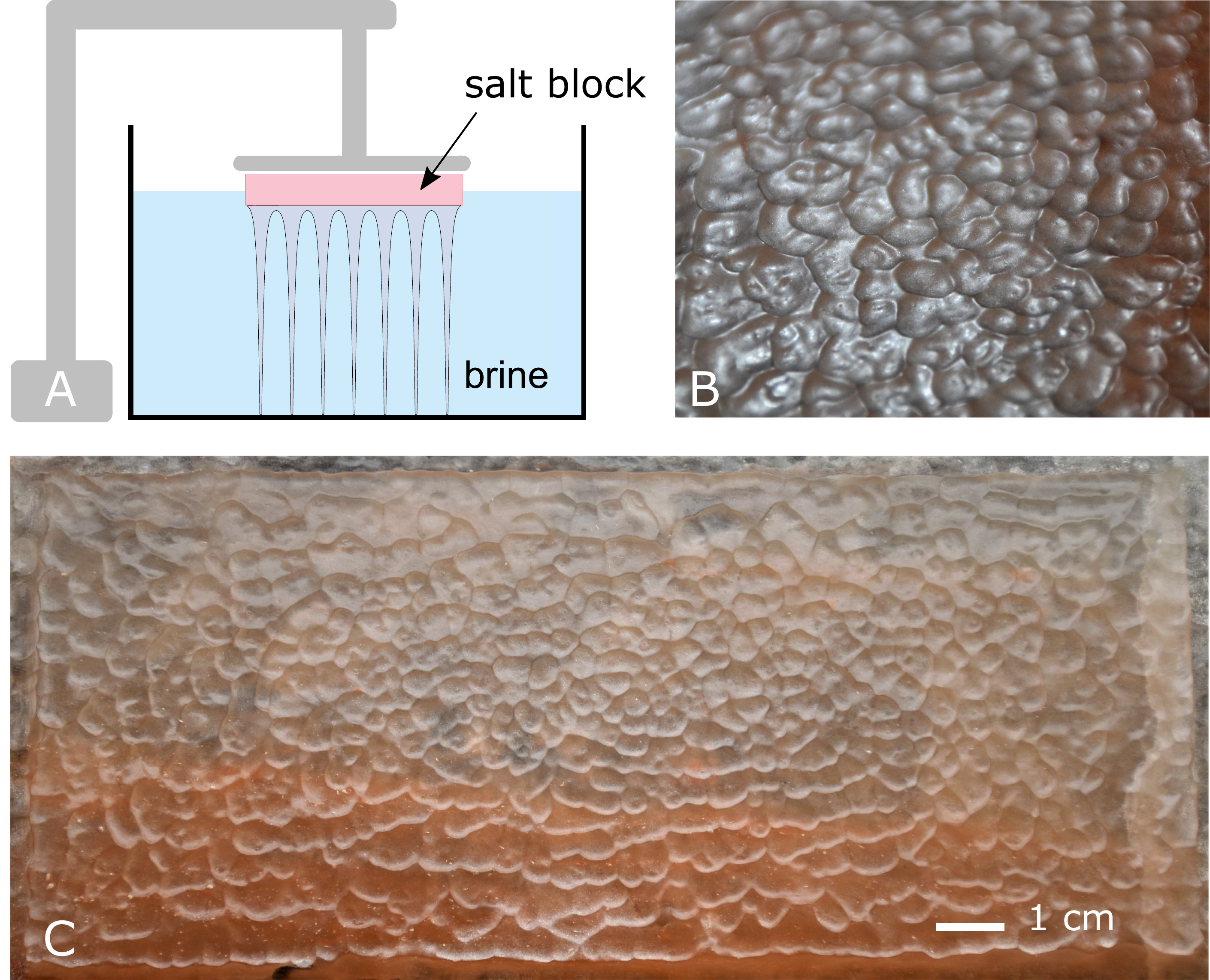}
\caption{Salt block experiment. A: Simplified schematic of the experimental set-up. A block of Himalayan pink salt of dimensions $\SI{20}{\centi\meter}\times\SI{10}{\centi\meter}\times\SI{2}{\centi\meter}$ is suspended in a brine of density $\rho_0=\SI{1.195}{\kilo\gram\per\liter}$. The concentration boundary layer at the bottom surface of the block is unstable and destabilizes into plumes which sink in the bottom of the tank. It induces a variable erosion rate, and thus leads to the emergence of a pattern. B: Photograph of a detail of the bottom surface of the salt block in the final state, after 220 hours of dissolution. A cellular pattern of scallops (concavities surrounded by sharp crests) has appeared. C: Photograph of the whole block in the final state. }
\label{fig:experience}
\end{figure*}

\end{document}